\documentstyle[11pt,nyas]{article}

\input psfig

\def\th{\thinspace}

\begin{document}

\title{Coherent Transport  of Angular Momentum\\
The Ranque--Hilsch Tube as a Paradigm}

\author{Stirling A. Colgate}
\affil{ Los Alamos National Lab \& New Mexico Tech}

\author{J. Robert Buchler}
\affil{University of  Florida}

\begin{abstract}
\textbf{ The mechanism for efficient and coherent angular momentum transport remains
one of the unsolved puzzles in astrophysics despite the enormous efforts that
have been made.  We suggest that important new insight could be gained in this
problem through an experimental and theoretical study of a laboratory device
(Ranque-Hilsch tube) that displays a similar enhanced angular momentum transfer
which cannot be explained by a simple turbulent model.  There is already good
experimental evidence to suggest that the cause of this enhancement is the
formation of aligned vortices that swirl around the symmetry axes very much
like virtual paddle blades.}
 \end{abstract}

\vskip 15pt


\section {Introduction}

It is not an exaggeration to say that the angular momentum transport is one of
the most important, yet poorest understood phenomena in astrophysics.
Furthermore, the angular momentum problem is ubiquitous, not only in the the
formation of the stars from the proto-stellar nebula, but in particular the Sun
and the planets, galaxies and their central black holes (including our own
Galaxy), and X ray sources powered by accretion from disks (e.g. Dubrulle 1993,
Papaloizou \& Lin 1995).  There is just too much initial angular momentum.  No
one has yet provided a full understanding of how it is transported outward so
fast and without concomitant excessive heating, as indicated by the
astronomical observations, and by our very existence on this planet.  Possible
exceptions involve situations with an external magnetic field and the necessary
ionization and conductivity (e.g. Hawley, Gammie \& Balbus 1995) which do not
exist in all situations where enhanced angular momentum is required.

Countless theories and many hundreds of theoretical papers in astrophysics have
sought this explanation, and then parameterize this lack of understanding by
the value of $\alpha$ as in the ubiquitous $\alpha$-viscosity (e.g.  Dubrulle
1993, Papaloizou \& Lin 1995).  This viscosity is orders of magnitude
greater than would be expected from laminar flow.  A time-independent, $k -
\epsilon$ turbulence model of the Ranque-Hilsch tube (see below) also shows a
need to scale up artificially the turbulent Prandtl number.  It is clear that
for a complete understanding of the angular momentum transport, full
time-dependent hydrodynamic modeling is needed.  However, it is even more
difficult than this for the astrophysicist, because gravity strongly stabilizes
the accretion disk, and the lack of a linear instability compounds the
difficulty of finding the mechanism that fuels turbulence and large-scale
structures such as vortices in the accretion disk.  One can even arbitrarily
introduce turbulence in a numerical model of an accretion disk (Balbus \&
Hawley, 1998, Hawley, Gammie \& Balbus, 1995) and observe it to rapidly decay
to laminar flow.

Coherent X-ray active structures have been reported on the surfaces of
accretion disks (Abramowicz et al. 1992), and in fact it has already been
proposed (Bath, Evans \& Papaloizou 1984) that short-term flickering of
cataclysmic variables and X-ray binaries have their origin in large-scale
vortices.

At Los Alamos, Lovelace, Li, Colgate, Nelson (1998), and Li, Finn, Colgate \&
Lovelace (1999) have derived analytically a linear growing, {\it azimuthally
nonsymmetric} instability in the disk provided one starts with a finite
initial radial entropy or pressure gradient. The Ranque-Hilsch tube is subject
to this same Rossby instability where a radial pressure gradient is induced by
the tangential injection of compressed air (or gas). The pressure
gradient induces a nonuniform distribution of vorticity or angular velocity,
which in turn is a sufficient condition for the the induction of the Rossby
instability.

The necessity for this nonuniform distribution of vorticity is shown in detail
for baroclinic flows by Staley \& Gall (1979). It is discussed by them in the
context of tornados, but the conditions are similar to the Ranque-Hilsch tube
where the radial pressure gradient presumably plays an identical role in
exciting the Rossby wave instability. The same criterion of a local maximum or
minimum of vorticity is necessary for the instability to occur in Keplerian
flows.  Thus one expects this instability to be the basis of the Ranque-Hilsch
tube and further expects that it is the nonlinear interaction of these
vortices that produces the weird effect of refrigeration.  Refrigeration is the
most dramatic experimental effect of the  Ranque-Hilsch tube. Later we will
offer an explanation of this effect in terms of these semi-coherent vortices.

 By observing, understanding, and modeling this phenomenon one will have a
laboratory example of the most likely mechanism of the enhanced transport of
angular momentum in a Keplerian accretion disk.

We review next some of the history of the Ranque-Hilsch tube.

\section{The Ranque -- Hilsch Tube}

 The Ranque-Hilsch tube (or vortex tube) was invented by Ranque (1933) and
improved by Hilsch (1946).  It is made up of a cylinder in which gas (air) is
injected tangentially, and at several atmospheres, through a nozzle of smaller
area than the tube, which sets up strong vortical flow in the tube.  Two exit
ports of comparable area to the nozzle allow the gas to escape, one port being
on axis and the second at the periphery.  The air flow is shown schematically
in Fig.~1.  The surprising result is that one stream is hot and the other
stream is cold. The question is which one is which and why.  Gas injected into
a static chamber of whatever form would in general be expected to exit through
any arbitrary ports and return to its original temperature when brought to
rest, assuming of course, that no heat is added or removed from the walls of
the chamber.  The contrary result for the Ranque-Hilsch tube has confounded
physicists and engineers for decades.  The device, when adjusted suitably, is
impressive with the hot side becoming too hot to touch and the cold side icing
up! However, the thermodynamic efficiency is poor, $\sim 20 \%$ to $30 \%$ of a
good mechanical refrigerator.

 \vskip 10pt
 \centerline{\psfig{figure=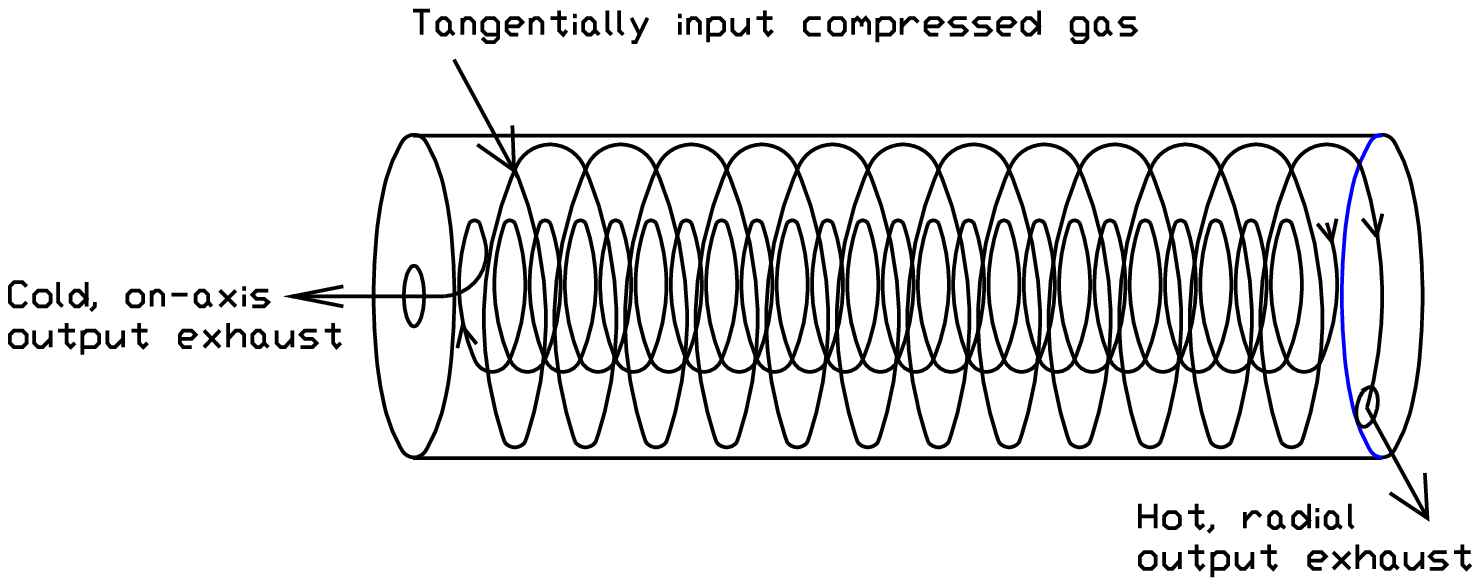,width=12cm}}
\centerline{{\small Fig.~1.~~Schematic of the Ranque-Hilsch tube}}
 \vskip 25pt

Vortex tubes are commercially available, both for practical applications (e.g.
cooling of firemen's suits) and for laboratory demonstrations. Completely
erroneous explanations are unfortunately frequently offered (for example, that
the tube separates the hot and cold molecules - a Maxwell demon! - clear in
violation of the second law of thermodynamics).

Previous experimental and theoretical work suggests that the Ranque-Hilsch tube
operates through the induction of co-rotating vortices in rotational flow.  The
reasons for this belief depend upon a review of the theory and measurements
described in the next section, but in summary these are: \th 1.)  the
unreasonableness of the high turbulent Prandtl number required to explain an
axisymmetric model; \th 2.)  the high frequency, large amplitude modulation
observed in pressure measurements and in the associated acoustic spectrum, \th 3.) the
reversibility of the temperature profile by entropy injection, \th 4.)  the
analogy to the temperature profile expected of a free running, radial flow
turbine.

We will review many of these theories and experiments later, but the point for
now is that there is no consensus on how this could happen, and to the extent
it happens, based upon current understanding of the solutions to the
Navier-Stokes equations.  The literature, both theory and experiment, has
recently been surveyed by Ahlborn \& Groves (1997) who conclude: ``This implies
that none of these mechanisms altogether explains the Ranque-Hilsch effect''.

As a recent numerical simulation of the Ranque-Hilsch tube with an axisymmetric
approximation (Fr\"ohlingsdorf \& Unger, 1999) shows that agreement with
observation requires the extraordinary value of the turbulent Prandtl number of
$\sim 9$ compared to unity for the $k - \epsilon$ turbulence model to obtain
agreement with the measured temperature difference of the two exit streams.  It
is precisely this very large departure from the "standard model" that makes
this device a paradigm for efficient and coherent angular momentum transport.
The vortex tube has an uncanny ability to efficiently transport angular
momentum and mechanical energy outward while severely limiting a
counterbalancing heat flow inward, a property shared by astrophysical accretion
disks.

It is fortunate that we have a laboratory device that can be used as a paradigm
for studying coherent transport of angular momentum.  It is true that the flow
field is extreme compared to that in astrophysical disks, but on the other hand
we do not have the complicating effects of gravity, nor of explaining the
origin of the vortices (instability and nonlinear growth), since they are
externally induced by the geometry of the tube.

Efficient turbine engines have very expensive blades that must withstand high
temperatures and stresses.  If the Navier-Stokes equations "know" a better way
of transporting angular momentum, then perhaps we could learn how to do so, and
to make a more efficient and cheaper engine.  We suggest that numerical
modeling combined with laboratory observations are the best way to find out and
improve our knowledge along the way.

\section{Angular Momentum and the Excitation of Rossby Vortices}

Let us first consider the flow in the cylinder under the assumption that the
flow is laminar. The Reynolds number in typical experiments is $ R_{e} \sim
10^5$, and so without turbulence, friction would be negligible with the
exception of the Ekman layer flow, to be considered later.  The primary
dynamical constraint under these circumstances is the conservation of angular
momentum, or $R v_{\phi} = R_{o} v_{\phi,o}$, so that the centrifugal
acceleration, with conserved angular momentum, becomes: $a = v_{\phi}^2/R =
(v_{o} R_{o})^2/ R^3$.  Since the centrifugal force must be balanced by
the pressure gradient,  we obtain

$$dP/dR = a \rho$$
 where $\rho$ is the gas density.  Using the adiabatic law

$$P = P_{o}(\rho / \rho_{o})^{\gamma}$$
 where $\gamma$  is the ratio of specific heats $= 1.4$ for air, and
integrating we have:


$$\left [ 1 - \bigg ({\rho\over\rho_o}\bigg)^{\gamma-1}
\right ] =
 Q \left [\bigg ({R_o\over R}\bigg )^2 -1 \right ]$$
where

$$Q =  {\gamma -1\over \gamma} {\rho_o\over P_o} {v_o^2\over 2} $$

Thus, in this approximation the density would vanish at finite radius unless
$Q$ is very small,  i.e. high input pressure compared to the input kinetic
energy.  Under normal operating conditions of the tube there would no way for
the gas to reach velocities sufficient to carry even a small fraction of the
input mass flow out the axial hole when one considers that $v_{o} \simeq (1/2)
\thinspace c_{s} $.  

Hence, the only way for the gas to exit the central port is to rid itself of
angular momentum as it spirals toward the axis.  Standard small scale
turbulence cannot do that without excessive concomitant heating.  Some large
scale eddy structure is needed, and our suggestion is that the flow has
non-azymuthally symmetric vorticity.

Two-dimensional rotational flow in an incompressible medium is known to be
unstable to the formation of "Rossby" waves (Nezlin \& Snezhkin, 1993).  The
general criterion for the instability is the existence of a local maximum or
minimum in an otherwise monotonic radial distribution of vorticity.  Such a
"bump" in vorticity is created by an entropy or pressure bump (c .f.  Fig. 1 of
Li et al. 1999).  When the waves grow to the nonlinear regime, they form
co-rotating vortices.  Such vortices act like particles in the sense that they
carry or transport a conserved mass in their cores.  Staley \& Gall (1979)
analyzed this instability and the structure of the vortices in the nonlinear
regime for tornados and found remarkable agreement between theoretical analysis
and the observations of 5 to 6 co-rotating vortices.  They conjectured, but
could not prove the enhanced transport of angular momentum by these vortices.
The latter are of particular interest in the atmospheric sciences because they
are responsible for most of the damage caused by tornados and hurricanes.  The
excitation of these vortices is also studied in planetary atmospheres where
theory (Marcus, 1988, 1990) and experiment (Sommeria, Meyers, \& Swinney, 1988)
demonstrate remarkable agreement with the observations of the "red spot" of the
Jovian atmosphere.  Multiple vortices can also be excited in laboratory
experiments where a thin layer of fluid is co-rotated in equilibrium within a
parabolic vessel into which vorticity is injected or removed at a local radius
(Nezlin \& Snezhkin, 1993), but again the question of the enhanced transport of
angular momentum in the fluid is not measured.

We expect that these same vortices must be induced in the Ranque-Hilsch tube,
because the flow at the innermost radius should be unstable because of the
steep gradients in density and temperature.  One of us, SAC, performed an
experiment to prove this at Lawrence Livermore National Lab as a basis for an
applied vortex reactor (Colgate, 1964).  Here a standard Ranque-Hilsch tube
produced the standard temperature ratios of a cold flow from the axial port and
a hot stream from the periphery. We suspected that if the instability was due
to the steep gradient in density and temperature, then a large change in the
entropy of the rotating gas stream at an intermediate radius would make a
significant change in the Ranque-Hilsch tube characteristics.  Consequently we
injected a flammable gas, acetylene, through a small hypodermic needle at a
flow rate close to stoichiometric at half radius of the tube. With ignition of
a flame, the radial temperature gradient was inverted by an order of magnitude,
and the typical Ranque-Hilsch tube exit temperature ratio was inverted.  The
axial exit stream became hot enough to melt tungsten, $\sim 4000$ deg and the
outer walls and peripheral exit stream returned to the input stream
temperature.  We thus became convinced that the the vortex flow field could be
stabilized by an entropy gradient, and the converse that the vortex flow
without a strong entropy gradient was unstable and that this instability was
fundamental to the refrigerator action of the Ranque-Hilsch tube.

The most reasonable explanation is that this instability induces axially
aligned vortices that act like semi-rigid vanes or turbine blades in
the flow.  These rigid members then transport mechanical work from the faster
rotating (higher vorticity) inner rotating flow to the periphery where
friction converts this mechanical energy to heat. The peripheral exit flow
then removes this heat. This is just what would be expected of a free running,
no mechanical load, radial flow turbine.  The rotor and blades would remove
mechanical energy from the axial exit stream and convert it into higher
velocity frictionally heated flow at the peripheral walls.

We would like to know whether and how these vortices can transport angular
momentum analogously to turbine blades. Before discussing these measurements we
need to take note of another source of enhanced frictional torque on the fluid,
namely the Ekman layers. Ekman layer flow is fluid in frictional contact with
the two stationary end walls of the tube. The consequential lack of rotation is
also a lack of centrifugal force, and so such a stationary fluid is highly
buoyant relative to the rotating flow. Hence a gas or fluid layer, close to the
end walls will flow rapidly towards the axis. The velocity of the radial flow
is limited by the same friction that has slowed its rotation and so an
equilibrium, steady state flow is reached, i.e. the Ekman layer.  The
thickness of this flow is typically $\simeq R_{o} \thinspace
R_{e}^{\scriptscriptstyle 1/2}$ and velocity $v_{Ekm} \simeq 1/2 v_{o}$ and so
the fractional loss due to the two Ekman layers of the two ends will be
$R_{e}^{\scriptscriptstyle -1/2} \simeq 3 \times 10^{\scriptscriptstyle -3}$, or small compared to the primary
angular momentum transport phenomenon sought.  However, whether this source of
frictional torque can be totally neglected, as in nearly all analyses, needs to
be checked more carefully.

\vskip 10pt

Because of the short time scales associated with the Ranque-Hilsch tube, it is
difficult to make measurements of the flow field. All experimental work has
therefore been limited to measuring time averaged quantities, thus necessarily
hiding the putative sub-vortices that we think are responsible for the
efficient transport of angular momentum and mechanical energy.

There are, however, a number of direct and indirect indications that
unsteadiness plays a major role in the dynamics of the Ranque-Hilsch tube, and
that the temperature separation is strongly influenced by 'fluctuations'.

1. A measurement of the acoustic spectrum (Ahlborn \& Groves 1997) show a
continuum stretching from 500Hz to above 25kHz, but with some very strong
features near 19kHz.  For this particular experiment this frequency corresponds
to about 1/5 of the angular period of the flow.  This is precisely what we
would expect from ($\sim$ 5) swirling vortices.

2. Measurements by Kurosaka (1982) show that the suppression of the vortex
whistle leads to a decrease of the energy separation.

3. Fr\"ohlingsdorf \& Unger (1999) who model the flow in a steady, axisymmetric
approximation with a $k - \epsilon$ turbulence model require an artificial
enhancement of the Prandtl number by a staggering factor of 10 to fudge the
unsteadiness.

We suggest here that the unsteady features are actually a set of 5 or 6
swirling vortices into which the unstable axisymmetric vortex has broken up.

An a priori 'obvious' measurement of the flow field with laser scattering from
suspended particles, e.g. smoke, does not work because the high centrifugal
forces prevent the seed particles from reaching the inner region.  The only
viable approach seems to be Schlieren photography with an ultrafast camera, and
correlated, (axis and radius), pressure measurements with small in situ probes
up to 10 MHz.

\vskip 10pt

The most sophisticated numerical approach that has been applied to the study of
the vortex tube (Fr\"ohlingsdorf \& Unger, 1999) is the industrial flow
software CFX.  In our opinion these calculations are inadequate in two ways:
First they assume azimuthal symmetry (about the axis of the tube) and second
they assume a steady flow.  Both of these assumptions hide the interesting
physics that we need to understand.  Proper 3D, or at least 2D numerical
simulations are possible, and are required to remedy these obvious
deficiencies.  The flow is expected to be subsonic, although we cannot rule out
shocks a priori and have to keep them in mind.  Molecular heat transport is
negligible.  The flow problem is thus mathematically speaking simple, in the
sense that it involves only the Navier Stokes equation.  However, Reynolds
numbers can be large, and the outer boundary layer plays an important
role. From a physical and numerical point of view one is thus faced with a
doable, although very difficult problem.

\section {Conclusion}

The Ranque-Hilsch tube has been a long standing
scientific puzzle since the first half of the century (Ranque 1933, Hilsch
1946), and many dozens of theoretical and experimental papers have been written
attempting to understand its action.  The frustration of lack of clear results
has often relegated the topic to "curiosity" status.  However, the equally
enigmatic and similar phenomenon in astrophysics of the Keplerian accretion
disk has been confronted in literally thousands of papers with similar results.
The observational fact of these high angular momentum transporting flows has
become indisputable so that the research effort has increased.  One result of
this effort has been to link theoretically the plausible explanation of the
Ranque-Hilsch tube to the Keplerian accretion disk.  The results have become
promising.  The expected excitation of axially aligned "Rossby" vortices has
been predicted analytically and the linear growth has been modeled with
numerical codes.  The excitation and action of these vortices represents new
physics.  One is reluctant to claim new physics without experimental proof.
One can obtain this proof in the laboratory and thereby complete hopefully the
last major non-understood domain of the Navier-Stokes equations.

\section {Acknowledgements}

This work has been aided by discussions with many colleagues among whom are Hui
Li, Richard Lovelace, David Ramond, Howard Beckley, and Van Romero.  It has
been partially supported by New Mexico Tech and by the Department of Energy
under contract w-7405-ENG-36, and by NSF (AST 9528338, and AST 9819608) at UF.

\end{document}